
\catcode`@=11

\newskip\ttglue

\font\twelverm=cmr12 \font\twelvebf=cmbx12
\font\twelveit=cmti12 \font\twelvesl=cmsl12

\font\ninerm=cmr9
\font\eightrm=cmr8
\font\sixrm=cmr6
\font\eighti=cmmi8   \skewchar\eighti='177
\font\sixi=cmmi6     \skewchar\sixi='177
\font\ninesy=cmsy9   \skewchar\ninesy='60
\font\eightsy=cmsy8  \skewchar\eightsy='60
\font\sixsy=cmsy6    \skewchar\sixsy='60
\font\eightbf=cmbx8
\font\sixbf=cmbx6
\font\eighttt=cmtt8  \hyphenchar\eighttt=-1
\font\eightit=cmti8
\font\eightsl=cmsl8

\def\smalltype{\def\rm{\fam0\eightrm}
 			\textfont0=\eightrm  \scriptfont0=\sixrm  \scriptscriptfont0=\fiverm
 			\textfont1=\eighti   \scriptfont1=\sixi   \scriptscriptfont1=\fivei
 			\textfont2=\eightsy  \scriptfont2=\sixsy  \scriptscriptfont2=\fivesy
 			\textfont3=\tenex    \scriptfont3=\tenex  \scriptscriptfont3=\tenex
    \textfont\itfam=\eightit  \def\it{\fam\itfam\eightit}
	   \textfont\slfam=\eightsl  \def\sl{\fam\slfam\eightsl}
	   \textfont\ttfam=\eighttt  \def\tt{\fam\ttfam\eighttt}
    \textfont\bffam=\eightbf  \scriptfont\bffam=\sixbf
        \scriptscriptfont\bffam=\fivebf  \def\bf{\fam\bffam\eightbf}
    \tt  \ttglue=.5em plus.25em minus.15em
    \normalbaselineskip=9pt
    \setbox\strutbox=\hbox{\vrule height7pt depth2pt width0pt}
    \let\sc=\sixrm  \let\big=\eightbig  \normalbaselines\rm}
\def\eightbig#1{{\hbox{$\textfont0=\ninerm\textfont2=\ninesy
    \left#1\vbox to6.5pt{}\right.\n@space$}}}

\def\medtype{\def\rm{\fam0\tenrm}
 			\textfont0=\tenrm  \scriptfont0=\sevenrm  \scriptscriptfont0=\fiverm
 			\textfont1=\teni   \scriptfont1=\seveni   \scriptscriptfont1=\fivei
 			\textfont2=\tensy  \scriptfont2=\sevensy  \scriptscriptfont2=\fivesy
 			\textfont3=\tenex    \scriptfont3=\tenex  \scriptscriptfont3=\tenex
    \textfont\itfam=\tenit  \def\it{\fam\itfam\tenit}
	   \textfont\slfam=\tensl  \def\sl{\fam\slfam\tensl}
	   \textfont\ttfam=\tentt  \def\tt{\fam\ttfam\tentt}
    \textfont\bffam=\tenbf  \scriptfont\bffam=\sevenbf
        \scriptscriptfont\bffam=\fivebf  \def\bf{\fam\bffam\tenbf}
    \tt  \ttglue=.5em plus.25em minus.15em
    \normalbaselineskip=12pt
    \setbox\strutbox=\hbox{\vrule height8.5pt depth3.5pt width0pt}
    \let\sc=\eightrm  \let\big=\tenbig  \normalbaselines\rm}

\def\bigtype{\let\rm=\twelverm \let\bf=\twelvebf
\let\it=\twelveit \let\sl=\twelvesl \rm}

\def\footnote#1{\edef\@sf{\spacefactor\the\spacefactor}#1\@sf
    \insert\footins\bgroup\smalltype
    \interlinepenalty100 \let\par=\endgraf
    \leftskip=0pt  \rightskip=0pt
    \splittopskip=10pt plus 1pt minus 1pt \floatingpenalty=20000
  \vskip4pt\noindent\hskip20pt\llap{#1\enspace}
\bgroup\strut\aftergroup\@foot\let\next}
\skip\footins=12pt plus 2pt minus 4pt \dimen\footins=30pc

\def\bigfont{\magnification=1200 \baselineskip=20pt}

 \def\b{\beta} 
  \def\k{\kappa}
\def\s{\sigma}

\def\cl#1{\centerline{#1}}
\def\clbf#1{\centerline{\bf #1}}

\def\is#1{{\narrower\smallskip\noindent#1\smallskip}}

\long\def\myname{\medskip
\cl{Kiho Yoon}
\cl{Department of Economics, Korea University}
\cl{145 Anam-ro, Seongbuk-gu, Seoul, Korea 02841}
\cl{ \tt kiho@korea.ac.kr}
\cl{\tt http://econ.korea.ac.kr/\~{ }kiho}
\medskip}

\def\ve{\vfill\eject}

\def\frac#1#2{{#1 \over #2}}
\def\Re{I\!\!R}

\newcount\sectnumber
\def\Section#1{\global\advance\sectnumber by 1 \bigskip
           \noindent{\bigtype {\bf \the\sectnumber  \ \ \ #1}} \medskip}

\def\defn#1{\medskip\noindent {\bf Definition #1. }}
\def\ass#1{\medskip\noindent {\bf Assumption #1.}}

\def\prop#1{\medskip\noindent {\bf Proposition #1.} \it}
\def\lemma#1{\medskip\noindent {\bf Lemma #1.} \it}
\def\thm#1{\medskip\noindent {\bf Theorem #1.} \it}
\def\corr#1{\medskip\noindent {\bf Corollary #1.} \it}

\def\eg#1{\medskip\noindent {\bf Example #1.}}

\def\ok{\smallskip \rm}

\def\pf{\medskip\noindent Proof: \/}

\def\endpf{\hfill {\it Q.E.D.} \smallskip}

\newcount\notenumber
\def\note#1{\global\advance\notenumber by 1
            \footnote{$^{\the\notenumber}$}{#1}\tenrm}

\def\ref{\bigskip \centerline{\bf REFERENCES} \medskip}

\def\emet{{\it Econometrica\/ }}
\def\jet{{\it Journal of Economic Theory\/ }}

\def\res{{\it Review of Economic Studies\/ }}
\def\geb{{\it Games and Economic Behavior\/ }}

\def\te{{\it Theoretical Economics\/ }}

\def\paper#1#2#3#4#5{\noindent\hangindent=20pt#1 (#2), ``#3,'' #4, #5.\par}


\bigfont

{ \ }

\vskip 1cm

{\bigtype
\clbf{Robust double auction mechanisms\footnote*{This work was supported by Basic Science Research Program through the National Research Foundation of Korea (NRF) funded by the Ministry of Education (2021R1$|$1A4A01059254) and by a Korea University Grant (k2209871).}
}

}

\vskip 1cm
\bigskip

\myname

\vskip 0.5cm

\clbf{Abstract}
\is{\baselineskip=12pt We study the robust double auction mechanisms, that is, the double auction mechanisms that satisfy dominant strategy incentive compatibility, ex-post individual rationality and ex-post budget balance. We first establish that the price in any robust mechanism does not depend on the valuations of the trading players. We next establish that, with a non-bossiness assumption, the price in any robust mechanism does not depend on players' valuations at all, whether trading or non-trading. Our main result is the characterization result that, with a non-bossy assumption along with other assumptions on the properties of the mechanism, the generalized posted mechanism in which a constant price is posted for each possible set of traders is the only robust double auction mechanism. We also show that, even without the non-bossiness assumption, it is quite difficult to find a reasonable robust double auction mechanism other than the generalized posted price mechanism.}
\smallskip

\is{\baselineskip=12pt Keywords: double auction, posted price, robust mechanism design, dominant strategy, budget balance}
\smallskip

\is{\baselineskip=12pt  JEL Classification: C72; D47; D82}

\ve

\Section{Introduction}

Double auctions are widely used in many two-sided market situations such as stock exchanges as well as commodity markets. In a typical double auction, (i) there are many sellers and many buyers who have private information about their true valuations for the good, (ii) they simultaneously submit their respective offers and bids to the mechanism (or the center, the mediator, the clearing house, the social planner, etc.), and  (iii) the mechanism determines the volume and the terms of trade according to a pre-determined rule. Double auctions approximate many other trading settings in the real world. They also have great theoretical value in the study of price formation under private information.

In this paper, we study the double auction mechanisms that satisfy dominant strategy incentive compatibility, ex-post individual rationality and ex-post budget balance. We call this family of double auction mechanisms as {\it robust\/} double auction mechanisms. A special instance of the robust double auction mechanisms is the robust bilateral trading mechanism for the environments in which there is one seller who initially owns one indivisible good and there is one buyer who is interested in obtaining the good. Hagerty and Rogerson (1987) and \v Copi\v c and Ponsat\'\i (2016) have shown that the only robust bilateral trading mechanism is the posted price mechanism, i.e., the mechanism in which a price is posted in advance and the seller and the buyer either trade at this price or do not trade at all. The purpose of this paper is to obtain a similar characterization for the general robust double auction mechanisms.

In the next section, we first establish that the price in any robust mechanism does not depend on the valuations of the {\it trading\/} players. This result in particular implies that a robust bilateral trading mechanism  is a posted price mechanism. In contrast, we show that a robust double auction mechanism may not be a posted price mechanism when there are more than two players.\note{Hence, Hagerty and Rogerson's conjecture (in page 97 of their paper) that the posted price mechanism may be the only robust mechanism in `$n$-person trading problems' is incorrect.} We next establish that, with a non-bossiness assumption, the price in any robust mechanism does not depend on players' valuations at all, whether trading or non-trading. Our main result is the characterization result for the general double auction environments that, with a non-bossy assumption along with other assumptions on the properties of the mechanism, the generalized posted mechanism in which a constant price is posted for each possible set of traders is the only robust double auction mechanism. Section 3 further characterizes robust double auction mechanisms. We show that, even without the non-bossiness assumption, it is quite difficult to find a value-respecting robust double auction mechanism other than the generalized posted price mechanism.\note{Value-respecting mechanisms are those mechanisms that select the traders according to the valuations. See Section 3 for details.} Section 4 concludes.

There is a large literature on double auction mechanisms: Representative works include Wilson (1985), Rustichini {\it et al.\/} (1994), Satterthwaite and Williams (2002), Cripps and Swinkels (2006) and Satterthwaite {\it et al.\/} (2014).\note{See also Kojima and Yamashita (2017) and the references therein for double auctions with interdependent values.} Most of the papers in this literature examine double auction mechanisms under the Bayesian incentive compatibility postulate. In contrast, the current paper focuses on robust double auction mechanisms. Robustness is desirable since the equilibrium behavior does not depend on the fine details of the environments, such as players' beliefs about each other. We note that the only other double auction mechanisms that satisfy dominant strategy incentive compatibility is McAfee's (1992) mechanism and Yoon's (2001) modified Vickrey double auction mechanism, but these mechanisms do not satisfy ex-post budget balance.

The papers similar in spirit but examining settings quite different from ours are Barber\`a and Jackson (1995) and Miyagawa (2001). The former studies classical exchange economies with many goods and shows that the only robust mechanism, under some additional conditions, is the fixed-proportion anonymous trading mechanism. The latter studies housing markets (i.e., trading situations in which each player initially owns one object to trade) with money and shows that the only robust mechanism, under some additional conditions, is the fixed-price core mechanism.

The research on robust mechanism design is not limited to the trading settings. Recently, Drexl and Kleiner (2015) and Shao and Zhou (2016) study the robust allocation problem of assigning one indivisible private object to one of two players, whereas Kuzmics and Steg (2017) study the robust public good provision problem. On a more general level, there is a voluminous literature concerning the common knowledge assumption and robust mechanism design after the pioneering work of Bergemann and Morris (2005).\note{See Bergemann and Morris (2013) for an excellent introduction.}

\Section{The main characterization}

There is a set ${\cal S} =\{1, \ldots, m\}$ of sellers and a set ${\cal B} =\{1, \ldots, n\}$ of buyers for a good. Each seller owns one indivisible unit of the good to sell, and each buyer wants to buy at most one unit of the good. Hence, there are $m$ units of the good available. Seller $i$'s privately known valuation for the good is denoted by $s_i$, and buyer $j$'s privately known valuation for the good is denoted by $b_j$. We assume that $s_i \in [0,1]$ for all $i \in {\cal S}$ as well as $b_j \in [0,1]$ for all $j \in {\cal B}$. Let $s=(s_1, \ldots, s_m)$ and $b=(b_1, \ldots, b_n)$. In addition, let $v=(s,b)$ and $V=[0,1]^{m+n}$. Hence, $V$ is the set of possible valuation profiles of the sellers and the buyers. We use the usual notation such as $v=(s_i, s_{-i}, b)=(s, b_j, b_{-j})$. We also use $v=(v_i, v_{-i})$ for $i \in {\cal S}$ and $v=(v_j, v_{-j})$ for $j \in {\cal B}$. That is, $(v_i, v_{-i})=(s_i, s_{-i}, b)$ for $i \in {\cal S}$ where $v_i=s_i$ and $v_{-i}=(s_{-i},b)$, and $(v_j, v_{-j})=(s, b_j, b_{-j})$ for $j \in {\cal B}$ where $v_j=b_j$ and $v_{-j}=(s, b_{-j})$.

Let $p_i: V \rightarrow [0,1]$ be the allocation rule for seller $i \in {\cal S}$, and let $q_j: V \rightarrow [0,1]$ be the allocation rule for buyer $j \in {\cal B}$. Both $p_i(v)$ and $q_j(v)$ denote the probability of trade. In addition, let $x_i: V \rightarrow \Re$ be the transfer rule for seller $i \in {\cal S}$ such that $x_i(v)$ is the receipt of the money seller $i$ gets, and let $y_j: V \rightarrow \Re$ be the transfer rule for buyer $j \in {\cal B}$ such that $y_j(v)$ is the payment of the money buyer $j$ makes. Note that $x_i(v)$ may be positive or negative: when it is negative, seller $i$ pays that amount. Likewise, $y_j(v)$ may be positive or negative: when it is negative, buyer $j$ receives that amount. The payoff of seller $i \in {\cal S}$  and buyer $j \in {\cal B}$ is denoted respectively by
$$u_i(v)=x_i(v)-p_i(v)s_i {\rm \ \ \  and \ \ \ } u_j(v)=q_j(v)b_j - y_j(v).$$

Let $p=(p_1, \ldots, p_m), q=(q_1, \ldots, q_n), x=(x_1, \ldots, x_m)$, and $y=(y_1, \ldots, y_n)$. The mechanism $(p,q,x,y)$ is said to be {\it dominant strategy incentive compatible\/} if
$$\eqalign{&u_i(s,b) \geq x_i(s'_i, s_{-i}, b)-p_i(s'_i, s_{-i},b)s_i \ \ \forall i \in {\cal S}, \forall s_i, \forall s'_i, \forall s_{-i}, \forall b; \cr
&u_j(s,b) \geq q_j(s, b'_j, b_{-j})b_j - y_j(s, b'_j, b_{-j}) \ \ \forall j \in {\cal B}, \forall b_j, \forall b'_j, \forall b_{-j}, \forall s}\eqno(IC)$$
and {\it ex-post individually rational\/} if
$$u_i(v) \geq 0 \ \ \forall i \in {\cal S} {\rm  \ \ and \ \ } u_j(v) \geq 0 \ \ \forall j \in {\cal B}, \ \ \forall v \in V. \eqno(IR)$$
The mechanism $(p,q,x,y)$ is said to be {\it ex-post budget balancing\/} if
$$\sum_{i \in {\cal S}} x_i(v) = \sum_{j \in {\cal B}} y_j(v) \ \ \forall v \in V \eqno (BB)$$
and {\it non-wasteful\/} if
$$\sum_{i \in {\cal S}} p_i(v) = \sum_{j \in {\cal B}} q_j(v) \ \ \forall v \in V. \eqno(NW)$$
Note well that we may have $\sum_{i \in {\cal S}} p_i(v) > \sum_{j \in {\cal B}} q_j(v)$ if the mechanism withholds/destroys some units. We exclude this possibility. We will call the mechanisms that satisfy $(IC)$, $(IR)$, $(BB)$ and $(NW)$ as {\it robust double auction mechanisms.\/}
We first have the following standard lemma.

\lemma1 The mechanism $(p,q,x,y)$ is dominant strategy incentive compatible (IC) if and only if

\item{(i)} $\forall i \in {\cal S}, \forall v_{-i}$: $p_i(v)$ is weakly decreasing in $v_i$.
\item{(ii)} $\forall j \in {\cal B}, \forall v_{-j}$: $q_j(v)$ is weakly increasing in $v_j$.
\item{(iii)} $\forall i \in {\cal S}, \forall v$: $u_i(v)=\int_{v_{i}}^{1} p_i(w, v_{-i}) dw + u_i(1, v_{-i})$.
\item{(iv)} $\forall j \in {\cal B}, \forall v$: $u_j(v)=\int_{0}^{v_{j}} q_j(w, v_{-j})dw + u_j(0, v_{-j})$.
\ok

\pf Omitted since it is standard. See Myerson (1981), Myerson and Satterthwaite (1983), etc. \endpf

We will restrict our attention to the deterministic allocation rules that take only the value zero or one. Thus, $p_i: V \rightarrow \{0,1\}$ for all $i \in {\cal S}$ and $q_j: V \rightarrow \{0,1\}$ for all $j \in {\cal B}$. Let $S(v) = \{i \in {\cal S} | p_i(v)=1\}$ denote the set of trading sellers at $v$, and let $B(v) = \{j \in {\cal B} | q_j(v)=1\}$ denote the set of trading buyers at $v$.

A possible property that the mechanism may satisfy is that all sellers face one identical price and all buyers face another identical price.

\ass1 (Common price) $\forall v \in V, \forall \{i, i'\} \subseteq S(v), \forall \{j, j'\} \subseteq B(v)$:

\cl{$x_i(v) = x_{i'}(v) {\rm \ and \ } y_j(v) = y_{j'}(v).$}
\ok

Hence, there are two prices, the seller price (ask price) of $\pi^s(v)$ and the buyer price (bid price) of $\pi^b(v)$. Of course, these two prices may be the same. Note that $\pi^s(v)$ and $\pi^b(v)$ may vary with $v$. This assumption is reasonable since all units of the good are identical and the players supply or demand at most one unit. Observe in particular that, though the famous VCG (Vickrey-Clarke-Groves) mechanisms may in general induce different transfers of money across players, there is only one seller price when each seller supplies one unit and only one buyer price when each buyer demands at most one unit. In fact, this assumption seems to be a prerequisite for any centralized trading institution. Example 1 below shows that we may pair one seller and one buyer and let the trade occur only between them. Then, each seller-buyer pair is a separated market and different prices across the pairs may emerge. Although we do not explicitly consider decentralized trading or resale possibilities in the current mechanism design approach, if the buyers face different prices, say, then it is conceivable that a buyer of the more expensive unit could find a new trade opportunity with a seller to the advantage of both.

Another possible property that the mechanism may satisfy is that the payoff of a player with the worst possible valuation is equal to zero: Observe that $u_i(v)$ is lowest when $v_i = 1$ and $u_j(v)$ is lowest when $v_j=0$ by Lemma 1(iii)-(iv).

\ass2 (Zero payoff for the worst type)

\cl{$\forall i \in {\cal S}, \forall v_{-i}: u_i(1,v_{-i})=0$ and $\forall j \in {\cal B}, \forall v_{-j}: u_j(0, v_{-j})=0$.} \ok

Note that the condition $u_j(0, v_{-j})=0$ for the buyer $j \in {\cal B}$ holds if and only if $y_j(0, v_{-j})=0$ holds. The latter condition is often assumed in the literature. Observe that $(IR)$ implies $y_j(0, v_{-j}) \leq 0$ whereas the no subsidy condition implies $y_j(0, v_{-j}) \geq 0$, leading us to $y_j(0, v_{-j})=0$ and hence $u_j(0, v_{-j})=0$.\note{The no subsidy condition is (i) $\forall j \in {\cal B}, \forall v: y_j(v) \geq 0$, and (ii) $\forall i \in {\cal S}, \forall v: x_i(v) \leq p_i(v)$.} The seller case is essentially symmetric. Observe that $(IR)$ implies $x_i(1,v_{-i}) \geq p_i(1, v_{-i})$ whereas the no subsidy condition implies $x_i(1,v_{-i}) \leq p_i(1, v_{-i})$, leading us to $u_i(1, v_{-i})=0$.

We now characterize some necessary conditions that a robust double auction mechanism must satisfy. We use the notation $|A|$ to denote the cardinality of an arbitrary set $A$, and the notation $v_{-T}$ to denote the valuation profile of the players not in the set $T \subseteq {\cal S} \cup {\cal B}$.

\prop1 Let the mechanism $(p,q,x,y)$ be a robust double auction mechanism. Under Assumption 1 of common price and Assumption 2 of zero payoff for the worst type, the mechanism $(p,q,x,y)$ must be the following form: for all $v \in V$ and $T(v) = S(v) \cup B(v)$,

\item{(i)} $|S(v)|=|B(v)|$;
\item{(ii)} $x_i(v) = y_j(v) = \pi(v_{-T(v)})$ for all $i \in S(v)$ and for all $j \in B(v)$;
\item{(iii)} $x_i(v)=y_j(v)=0$ for all $i \notin S(v)$ and for all $j \notin B(v)$;
\item{(iv)} $v_i \leq \pi(v_{-T(v)})$ for all $i \in S(v)$ and $v_j \geq \pi(v_{-T(v)})$ for all $j \in B(v)$. \ok

\noindent {\sl Remark:} Property (i) says that the number of trading sellers must be equal to the number of trading buyers. Property (ii) says that all trading sellers and trading buyers face the same price $\pi(v_{-T(v)})$ that is independent of their valuations. Note in particular that there is only one price, i.e., the seller price is identical to the buyer price. Property (iii) says that the sellers and the buyers who do not trade neither receive nor pay any money. Property (iv) says that every trading seller's valuation is less than or equal to the price and every trading buyer's valuation is greater than or equal to the price.

\pf It is obvious that $(NW)$ implies property (i). Next, since $p_i(v_i, v_{-i}) \in \{0,1\}$, Lemma 1(i) implies: for a given $v_{-i}$, there exists $z_i(v_{-i}) \in [0, 1]$ such that
$$p_i(v_i, v_{-i}) = \cases{1 &if $v_i < z_i(v_{-i})$; \cr
                            0 &if $v_i > z_i(v_{-i})$.}$$
Likewise, Lemma 1(ii) implies: for a given $v_{-j}$, there exists $z_j(v_{-j}) \in [0, 1]$ such that
$$q_j(v_j, v_{-j}) = \cases{0 &if $v_j < z_j(v_{-j})$; \cr
                            1 &if $v_j > z_j(v_{-j})$.}$$
That is,
$$\eqalign{&z_i(v_{-i})=\sup \{v_i \in [0,1] | p_i(v_i, v_{-i}) = 1\} {\rm \ \ for \ } i \in {\cal S} ; \cr
&z_j(v_{-j})= \inf \{v_j \in [0, 1] | q_j(v_j, v_{-j}) = 1\} {\rm \ \ for \ } j \in {\cal B}.}$$
Then, by Lemma 1(iii) and Assumption 2,
$$u_i(v) = \cases{z_i(v_{-i}) - v_i &when $p_i(v_i, v_{-i})=1$; \cr
                  0 &when $p_i(v_i, v_{-i})=0$.}$$
Likewise, by Lemma 1(iv) and Assumption 2,
$$u_j(v)=\cases{v_j - z_j(v_{-j}) &when $q_j(v_j, v_{-j})=1$; \cr
                0 &when $q_j(v_j, v_{-j})=0$.}$$
Hence, we have:

\item{(i)} For a seller $i \in {\cal S}$:
$$x_i(v)=\cases{z_i(v_{-i}) &when $p_i(v)=1$; \cr
                0 &otherwise.}$$
\item{(ii)} For a buyer $j \in {\cal B}$:
$$y_j(v)=\cases{z_j(v_{-j}) &when $q_j(v)=1$; \cr
               0 &otherwise.}$$

We then have
$$\sum_{i \in S(v)} x_i(v) = \sum_{j \in B(v)} y_j(v) \ \ \forall v \in V$$
by $(BB)$ and the fact that $x_i(v)=0$ for all $i \notin S(v)$ and $y_j(v)=0$ for all $j \notin B(v)$. Since $\sum_{i \in S(v)} x_i(v) = |S(v)| \pi^s(v)$ and $\sum_{j \in B(v)} y_j(v) = |B(v)| \pi^b(v)$ by Assumption 1 and $|S(v)|=|B(v)|$ by (i), we have $\pi^s(v)=\pi^b(v)$. Let us denote this price by $\pi(v)$.

If $T(v) \ne \emptyset$, then $\pi(v)=z_k(v_{-k})$ for all $k \in T(v)$. Hence, $\pi(v)$ does not depend on $v_k$ for any $k \in T(v)$. So, we can write this price as $\pi(v_{-T(v)})$ for all $v \in V$ and $T(v) \subseteq {\cal S} \cup {\cal B}$. This proves that properties (ii) and (iii) hold. It is now clear that $(IR)$ implies property (iv). \endpf

Both Assumption 1 and Assumption 2 are essential for Proposition 1 as the following examples demonstrate. Note that Assumption 2 is satisfied in Example 1 whereas Assumption 1 is (trivially) satisfied in Example 2.

\eg1 Suppose there are two sellers and two buyers, and consider a mechanism in which (i) two distinct constant prices $\pi^s_1=\pi^b_1$ and $\pi^s_2=\pi^b_2$ are posted in advance, and (ii) a trade occurs between seller 1 and buyer 1 when and only when $s_1 < \pi^s_1=\pi^b_1 < b_1$ and a trade occurs between seller 2 and buyer 2 when and only when $s_2 < \pi^s_2=\pi^b_2 < b_2$. Hence, seller 1 and buyer 1 face the posted price $\pi^s_1=\pi^b_1$ and trade only between them, and likewise for seller 2 and buyer 2. This mechanism is a robust double auction mechanism that violates Assumption 1. In this example, there are two posted price mechanisms separately applied to two pairs of traders, the pair of seller 1 and buyer 1 and the pair of seller 2 and buyer 2.

\eg2 Suppose there are one seller and two buyers, and consider a mechanism in which (i) two prices, the seller price $\pi^s$ and the buyer price $\pi^b$, are posted in advance with $\pi^s < \pi^b$, (ii) a trade occurs between seller 1 and buyer 1 when and only when $s_1 < \pi^s < \pi^b < b_1$, (iii) when a trade occurs, seller 1 receives $\pi^s$, buyer 1 pays $\pi^b$, and buyer 2 receives $\pi^b - \pi^s$, that is, $x_1(v)=\pi^s$, $y_1(v)=\pi^b$, and $y_2(v)=\pi^s - \pi^b$,  and (iv) when there is no trade, every player's monetary transfer is equal to zero, that is, $x_1(v)=y_1(v)=y_2(v)=0$. This mechanism is a robust double auction mechanism without one price for all traders. The reason Proposition 1 does not hold in this example is that Assumption 2 is not satisfied.\note{We note that we can construct a similar mechanism for the case when there are more than one seller and more than one buyer.} \ok

Observe that, for bilateral trading with one seller and one buyer, i.e., ${\cal S}={\cal B}=\{1\}$, we always have $T(v)={\cal S} \cup {\cal B}$ unless $T(v)=\emptyset$. Hence, $\pi(v)$ does not depend on either $s_1$ or $b_1$ whenever $T(v) \ne \emptyset$. This mechanism can be implemented by a posted price mechanism: A constant price $\pi$ is posted in advance and the trade occurs when and only when $s_1 < \pi < b_1$.\note{We assume that players decide not to trade when they are indifferent between trade and no trade. We follow this convention throughout the paper.} Since it is straightforward to show that Assumptions 1 and 2 are satisfied for bilateral trading environments, Proposition 1 implies one of Hagerty and Rogerson's (1987) results that a robust double auction mechanism for bilateral trading environments is a posted price mechanism.

\corr{\hskip -4pt} A  double auction mechanism for bilateral trading environments is robust if and only if it is a posted price mechanism. \ok

\pf Observe that Assumption 1 is trivially satisfied for bilateral trading environments. Observe next that $x_1(s_1,b_1)=y_1(s_1,b_1)$ by $(BB)$ and $p_1(s_1, b_1)=q_1(s_1,b_1)$ by $(NW)$. Hence, the total surplus, i.e., the sum of players' payoffs, is $p_1(s_1, b_1)(b_1 - s_1)$. This becomes (i) $p_1(1, b_1)(b_1 -1) \leq 0$ when $s_1 = 1$, and (ii) $-p_1(s_1, 0) s_1 \leq 0$ when $b_1 = 0$. Since the total surplus is nonpositive when either $s_1=1$ or $b_1=0$ whereas each player's payoff is nonnegative by $(IR)$, the only possibility is that each player gets a zero payoff. Thus, Assumption 2 is always satisfied for bilateral trading environments. Proposition 1 then implies that a mechanism in these environments is robust only if it is a posted price mechanism. The converse also holds since a posted price mechanism is clearly robust. \endpf

In contrast, a robust double auction mechanism may not be a posted price mechanism when there are more than two players, as the following example demonstrates.

\eg3 Suppose there are one seller and two buyers, and consider a mechanism in which (i) players first submit their valuations (which may or may not be true valuations), (ii) the price is set to the lower of the buyers' reported valuations, and (iii) a trade occurs between the seller and the buyer of the higher valuation when and only when the seller's valuation is less than the price. When $s_1=0.3$, $b_1=0.7$, and $b_2 \in (0.3, 0.7)$, for instance, the trade occurs and the price $\pi = b_2$ depends on buyer 2's valuation. \ok

Observe in particular that the mechanism in this example is dominant strategy incentive compatible $(IC)$, mainly due to the fact that the traders cannot affect the terms of trade. On the other hand, the non-trading buyer affects the terms of trade without altering her own allocation or payment. We now introduce another property often used in the literature to obtain a sharper result. The mechanism $(p,q,x,y)$ is said to be {\it non-bossy\/} if
\item{(i)} For all $i \in {\cal S}$ and for all $v_i, v'_i$, and $v_{-i}$: if $p_i(v_i, v_{-i})=p_i(v'_i, v_{-i})$ and $x_i(v_i, v_{-i})=x_i(v'_i, v_{-i})$, then $x(v_i, v_{-i})=x(v'_i, v_{-i})$ and $y(v_i, v_{-i})=y(v'_i, v_{-i})$.
\item{(ii)} For all $j \in {\cal B}$ and for all $v_j, v'_j$, and $v_{-j}$: if $q_j(v_j, v_{-j})=q_j(v'_j, v_{-j})$ and $y_j(v_j, v_{-j})=y_j(v'_j, v_{-j})$, then $x(v_j, v_{-j})=x(v'_j, v_{-j})$ and $y(v_j, v_{-j})=y(v'_j, v_{-j})$.
\smallskip

\noindent In words, a mechanism is non-bossy if no player can change others' payments without changing his/her allocation or payment. We note that this definition is somewhat different from the standard one first introduced by Satterthwaite and Sonnenschein (1981). The standard non-bossiness property is stated as follows in the current setup: the mechanism $(p,q,x,y)$ is non-bossy if (i) for all $i \in {\cal S}$ and for all $v_i, v'_i$, and $v_{-i}$: if $p_i(v_i, v_{-i})=p_i(v'_i, v_{-i})$ and $x_i(v_i, v_{-i})=x_i(v'_i, v_{-i})$, then $p(v_i, v_{-i})=p(v'_i, v_{-i})$, $q(v_i, v_{-i})=q(v'_i, v_{-i})$, $x(v_i, v_{-i})=x(v'_i, v_{-i})$ and $y(v_i, v_{-i})=y(v'_i, v_{-i})$; (ii) for all $j \in {\cal B}$ and for all $v_j, v'_j$, and $v_{-j}$: if $q_j(v_j, v_{-j})=q_j(v'_j, v_{-j})$ and $y_j(v_j, v_{-j})=y_j(v'_j, v_{-j})$, then $p(v_j, v_{-j})=p(v'_j, v_{-j})$, $q(v_j, v_{-j})=q(v'_j, v_{-j})$, $x(v_j, v_{-j})=x(v'_j, $ $v_{-j})$ and $y(v_j, v_{-j})=y(v'_j, v_{-j})$. That is, the conclusion part includes the statements about allocations as well as payments.\note{See Thomson (2016) for a thorough discussion of non-bossiness.} Hence, our definition is weaker than the standard definition.

\ass3 (Non-bossiness) The mechanism $(p,q,x,y)$ is non-bossy. \ok

The non-bossiness assumption is an extra property imposed on the mechanism on top of the properties required by Proposition 1 under which the generalized posted price mechanism results. We note that this assumption is also imposed (along with additional assumptions) in the works of Barber\`a and Jackson (1995) and Miyagawa (2001) cited in the introduction. We have:

\prop2 Let the mechanism $(p,q,x,y)$ be a robust double auction mechanism. Under Assumption 1 of common price, Assumption 2 of zero payoff for the worst type and Assumption 3 of non-bossiness, the mechanism $(p,q,x,y)$ must be the following form: for all $v \in V$ and $T(v) = S(v) \cup B(v)$, there exists a constant price $\pi$ such that

\item{(i)} $|S(v)|=|B(v)|$;
\item{(ii)} $x_i(v) = y_j(v) = \pi$ for all $i \in S(v)$ and for all $j \in B(v)$;
\item{(iii)} $x_i(v)=y_j(v)=0$ for all $i \notin S(v)$ and for all $j \notin B(v)$;
\item{(iv)} $v_i \leq \pi$ for all $i \in S(v)$ and $v_j \geq \pi$ for all $j \in B(v).$ \ok

\pf We know by Proposition 1(ii) that the price does not depend on the valuation profile of the traders, i.e., $\pi(v)=\pi(v_{-T(v)})$ for all $v \in V$. We now show that it does not depend on the valuation profile of the non-traders, either. Let there be a non-trading seller $i \in S^c(v_i, v_{-i}) \cap S^c(v'_i, v_{-i})$. This means $p_i(v_i, v_{-i}) = 0 = p_i(v'_i, v_{-i})$, and $x_i(v_i, v_{-i}) = 0 = x_i(v'_i, v_{-i})$ by Proposition 1(iii). We have $\pi(v_i, v_{-T(v) \setminus \{i\}})= \pi(v_i, v_{-i}) = \pi(v'_i, v_{-i}) = \pi(v'_i, v_{-T(v) \setminus \{i\}})$ where the second inequality holds by Assumption 3. The argument for a non-trading buyer is similar. \endpf

\noindent Hence, the set $T(v) = S(v) \cup B(v)$ of traders may vary as the valuation profile $v$ changes, but the price does not vary for given a set of traders. We note that this constant price may be different for different set of traders, as the following example demonstrates.

\eg4 Suppose there are one seller and two buyers. Let $\pi$ and $\pi'$ be two constant prices with $\pi > \pi'$. Let the mechanism be such that
$$\eqalign{&(q_1(v), y_1(v)) = \cases{(1, \pi) & if $s_1 < \pi < b_1$; \cr
                                      (0, 0) & otherwise,} \cr
           &(q_2(v), y_2(v)) = \cases{(1, \pi') & if $s_1 < \pi' < b_2$ but not $s_1 < \pi < b_1$; \cr
                                      (0, 0) & otherwise,} \cr
           &(p_1(v), x_1(v)) = (q_1(v) + q_2(v), y_1(v) + y_2(v)).
}$$
It is straightforward to show that this mechanism satisfies all properties in Proposition 2.

Since both the set of sellers and the set of buyers are finite, the family of the sets of traders is also finite. This leads us to the following definition.

\defn{\hskip -5pt} A mechanism is a generalized posted price mechanism if, for each set $T = S \cup B$ of traders with $|S| = |B|$, there is a constant price $\pi$ such that all sellers in $S$ receive $\pi$, all buyers in $B$ pay $\pi$, and all other players receive or pay nothing. \ok

Proposition 2 thus states that the price is posted for each set of traders in any robust double auction mechanism that also satisfies Assumptions 1, 2, and 3. Restating it, we have:

\thm{\hskip -5pt}  A robust double auction mechanism that also satisfies Assumption 1 of common price, Assumption 2 of zero payoff for the worst type and Assumption 3 of non-bossiness must be a generalized posted price mechanism. \ok

We note that the converse of the theorem is not true. To see this, consider Example 4 again but with $\pi < \pi'$. The seller with $s_1 < \pi$ has an incentive to submit $s'_1$ with $\pi \leq s'_1 < \pi'$ and trade with buyer 2 when $\pi < b_1$ and $\pi' < b_2$. Thus, this generalized posted price mechanism is not dominant strategy incentive compatible. We also note that the theorem is a statement about the terms of trade, but not about the selection of traders. There may exist many alternative manners of determining the set $T(v) = S(v) \cup B(v)$ of traders for each possible $v$, and it appears to be a difficult task to provide a clear characterization.

\Section{The linear price mechanism}

Are there robust double auction mechanisms that satisfy Assumptions 1 and 2 but not Assumption 3? That is, are there robust double auction mechanisms other than the generalized posted-price mechanisms? We already know by Example 3 that there are such mechanisms. In this section, we further pursue this question. We introduce a canonical double auction mechanism, the linear price mechanism, and analyze it to see what might happen when we dispense with the non-bossiness property.

Let us first consider the following variant of Example 3. Suppose there are one seller and two buyers, and consider a mechanism in which (i) the price is set to buyer 2's reported valuation $b_2$, and (ii) a trade occurs between seller 1 and buyer 1 when and only when $s_1 < b_2 < b_1$. In this example, the only possible trade is between seller 1 and buyer 1, and buyer 2 contributes to the determination of the price. It is easy to see that this is a robust double auction mechanism. However, this mechanism has the undesirable feature that buyer 2 never has an opportunity to trade even when her valuation is high.

This leads us to consider those mechanisms that respect all players' valuations, which we will call the {\it value-respecting\/} mechanisms. If the selection of traders depends on players' valuations, then it must be the case that the sellers are selected in an increasing of valuations and the buyers are selected in a decreasing order of valuations. That is, sellers with lower valuations and buyers with higher valuations are given priorities. Otherwise, a seller (buyer) who does not trade but has a lower (higher) valuation than a seller (buyer) who trades may have an incentive to bid higher (lower) to match the latter's valuation, destroying incentive compatibility. Moreover, a robust double auction mechanism satisfying Assumptions 1 and 2 (but not Assumption 3) must have the necessary properties derived in Proposition 1. In particular, the price does not depend on the players' valuations in $T(v)=S(v) \cup B(v)$. The following mechanism is a natural mechanism with those properties.
\ve

Players first submit their respective valuations in this mechanism, which may or may not be true valuations. Let us order sellers' submitted valuations in an increasing order and buyers' submitted valuations in a decreasing order. Thus, we have the order statistics $s_{(1)} \leq s_{(2)} \leq \cdots \leq s_{(m)}$ of sellers' valuations and the order statistics $b_{(1)} \geq b_{(2)} \geq \cdots \geq b_{(n)}$ of buyers' valuations. Ties in the valuations can be broken in any predetermined way. For convenience, we follow the convention that $s_{(0)}=b_{(n+1)}=0$ and $s_{(m+1)}=b_{(0)}=1$. Let $\k$ be the number such that $s_{(\k)} \leq b_{(\k)}$ and $s_{(\k+1)} > b_{(\k+1)}$. That is, $\k = \max \{k | s_{(k)} \leq b_{(k)} \}$. Let us call $\k$ as the tentative volume of trade. When $\k < \min\{m,n\}$, the posted price $\pi$ is set to
$$\pi = c_{1} s_{(\k+1)} + c_{2} s_{(\k+2)} +\cdots + c_{m-\k} s_{(m)}+d_{1} b_{(\k+1)}+ d_{2} b_{(\k+2)}+\cdots + d_{n-\k} b_{(n)}$$
for given nonnegative numbers $c_{1}, \ldots, c_m, d_{1}, \ldots, d_n$.\note{Any pre-specified rule for the case when $\k = \min\{m,n\}$ is fine for our purpose, so we do not specify it.} We assume these coefficients are such that $0 \leq \pi \leq 1$ holds for any realization of valuations, since the possibility of trade is precluded otherwise. In particular, since all valuations belong to the interval $[0,1]$, any convex combination of non-traders' valuations will induce $\pi$ between zero and one. Let $\hat S(v) = \{i \in {\cal S} | s_i \leq s_{(\k)}$ and $s_i < \pi \}$ and $\hat B(v) = \{j \in {\cal B} | \ b_j \geq b_{(\k)}$ and $b_j > \pi \}$. If $|\hat S(v)| = |\hat B(v)|$, then all traders in the set $\hat S(v)$ and $\hat B(v)$ trade at price $\pi$. Otherwise, rationing is needed. Let $\s: {\cal S} \rightarrow {\cal S}$ be a permutation of sellers such that seller $\s(1)$ is ordered first, seller $\s(2)$ is ordered second, and so on. Likewise, let $\b: {\cal B} \rightarrow {\cal B}$ be a permutation of buyers such that buyer $\b(1)$ is ordered first, buyer $\b(2)$ is ordered second, and so on. If $|\hat S(v)| > |\hat B(v)|$, then $|\hat B(v)|$ sellers in $\hat S(v)$ are selected in the order of $\s: \cal S \rightarrow \cal S$ to trade with the buyers in $\hat B(v)$; if $|\hat S(v)| < |\hat B(v)|$, then $|\hat S(v)|$ buyers in $\hat B(v)$ are selected in the order of $\b: \cal B \rightarrow \cal B$ to trade with the sellers in $\hat S(v)$.\note{If we allow probabilistic allocation rules, then random rationing is the most prominent rationing rule both in reality and in the literature: If $|\hat S(v)| > |\hat B(v)|$, then $|\hat B(v)|$ sellers in $\hat S(v)$ are selected randomly to trade with the buyers in $\hat B(v)$; if $|\hat S(v)| < |\hat B(v)|$, then $|\hat S(v)|$ buyers in $\hat B(v)$ are selected randomly to trade with the sellers in $\hat S(v)$. It is straightforward to see that Proposition 3 below can be extended with minor modification to the random rationing rule.}

Let us call this mechanism as the {\it linear price mechanism\/}. Observe that (i) the tentative volume of trade is set to maximize potential gains from trade, (ii) the price is a linear combination of non-traders' valuations, (iii) a seller (a buyer, respectively) may not trade even when $s_i \leq s_{(\k)}$ ($b_j \geq b_{(\k)}$, respectively) depending on the level of the price $\pi$ actually realized, (iv) the actual volume of trade is less than or equal to the tentative volume of trade $\k$, and (v) the players are obliged to follow the trading decision after the submission of valuations. The following example illustrates how a linear price mechanism works.

\eg{5} The price $\pi$ is set to $b_{(\k+1)}$. That is, $d_1=1$ and $c_1=\cdots=c_m=d_2=\cdots=d_n=0$. There are 4 sellers and 4 buyers. Let both $\s: \cal S \rightarrow \cal S$ and $\b: \cal B \rightarrow \cal B$ be identity mappings. When $s_1=0.1, s_2=0.3, s_3=0.5$, $s_4=0.7$, $b_1=0.9, b_2=0.8, b_3=0.6$ and $b_4=0.4$, we have $\k=3$, $\pi=0.4$, $\hat S(v)=\{1,2\}$ and $\hat B(v)=\{1,2,3\}$. Since $|\hat S(v)| < |\hat B(v)|$, the buyer side has to be rationed. Hence, both sellers 1 and 2 and buyers 1 and 2 trade. Note that both seller 3 and buyer 3 do not trade in spite of $\k=3$: Seller 3 does not trade since the price $\pi$ is lower than his valuation whereas buyer 3 does not trade since she is ordered after buyers 1 and 2 in the rationing rule $\b$.

We now show that, when there are at least two sellers as well as at least two buyers, this mechanism does not satisfy $(IC)$. We start with the following lemma.

\lemma2 Let $\min \{m,n\} \geq 2$. If a linear price mechanism satisfies $(IC)$, then
$$c_{1}=c_{2}=\cdots=c_{m}=0.$$ \ok

\pf We first show that $c_{1}=0$. Suppose to the effect of contradiction that $c_{1} > 0$. Let the realized valuations be such that $s_{(\k)} < b_{(\k+1)} < s_{(\k+1)} < c_{1} s_{(\k+2)}$ and $b_{(1)}$ is sufficiently high.\note{It helps to think of the case when $\k=1$.} Then, seller $i$ with $s_i = s_{(\k+1)}$ has an incentive to bid arbitrarily close to 0 when $\s(1)=i$. In that case, the new tentative volume of trade becomes $\k+1$ and the new price becomes $\pi' \geq c_{1} s_{(\k+2)}$, which is higher than $s_{(\k+1)}$. Since there exists at least one buyer whose valuation $b_{(1)}$ is higher than $\pi'$, seller $i$ gets a positive payoff instead of a payoff of zero from reporting truthfully.

We next show that $c_l=0$ for all $l = 2, \ldots, m$. Suppose to the effect of contradiction that $c_l > 0$. Let the realized valuations be such that $b_{(\k+1)} < s_{(\k)} < s_{(\k+1)} < \pi$ and $b_{(1)}$ is sufficiently high. For instance, let $s_{(\k+1)}$ be small enough to satisfy $s_{(\k+1)} < c_l s_{(\k+l)}$. Then, seller $i$ with $s_i = s_{(\k+1)}$ has an incentive to bid arbitrarily close to zero when $\s(1)=i$ since (i) it would increase the probability of trade from zero to one as well as (ii) it would not change the price since the new price\note{Observe that the tentative volume of trade does not change. Observe also that, although $s_{(\k)}$ now becomes the $(\k+1)$-th lowest valuation of the seller, the price remains the same since $c_1 = 0$.}
$$\pi' = c_{2} s_{(\k+2)} +\cdots + c_{m-\k} s_{(m)}+d_{1} b_{(\k+1)}+ d_{2} b_{(\k+2)}+\cdots + d_{n-\k} b_{(n)}$$
is equal to the original price $\pi$, from which he gets a positive payoff instead of a payoff of zero from reporting truthfully.  \endpf

Hence, we have $\pi=d_{1} b_{(\k+1)} + \cdots + d_{n-\k} b_{(n)}$. We now show that this mechanism does not satisfy $(IC)$. Let the realized valuations be such that \par
\cl{$d_{1} b_{(\k+2)} + \cdots + d_{n-\k-1} b_{(n)} < b_{(\k+1)} < s_{(\k+1)} < b_{(\k)}$}
\noindent and $s_{(1)}$ is sufficiently low. Then, buyer $j$ with $b_j = b_{(\k+1)}$ has an incentive to bid arbitrarily close to 1 when $\b(1)=j$. In that case, the new tentative volume of trade becomes $\k+1$ and the new price becomes $\pi'=d_{1} b_{(\k+2)} + \cdots + d_{n-\k-1} b_{(n)}$, which is lower than $b_{(\k+1)}$. Since there exists at least one seller whose valuation $s_{(1)}$ is lower than $\pi'$, buyer $j$ gets a positive payoff instead of a payoff of zero from reporting truthfully. Summarizing the discussion, we have:

\prop3 The linear price mechanism is not a robust double auction mechanism when there are more than one seller and more than one buyer. \ok

\noindent Hence, even without Assumption 3 of non-bossiness, it seems quite difficult to find a value-respecting robust double auction mechanism other than the generalized posted price mechanism.

\Section{Conclusion}

We have shown that the price in any robust double auction mechanism does not depend on the valuations of the trading sellers or the trading buyers. We have also shown that, with a non-bossiness assumption, the price in any robust double auction mechanism does not depend on the players' valuations at all, whether trading or non-trading. Our main result is the characterization result for the general double auction environments that, with a non-bossy assumption along with other assumptions on the properties of the mechanism, the generalized posted mechanism is the only robust double auction mechanism. Furthermore, we have introduced the linear price mechanism to demonstrate that, even without the non-bossiness assumption, it is quite difficult to find a reasonable robust double auction mechanism other than the generalized posted price mechanisms.

Although some of the results can be extended straightforwardly to the class of probabilistic allocation rules, our main characterization has been established for deterministic allocation rules. Hence, it remains as a future research agenda to extend the analysis to probabilistic robust double auction mechanisms.

\ref

\paper{Barber\`a, S., Jackson, M.}{1995}{Strategy-proof exchange}{\emet 63}{51-87}

\paper{Bergemann, D., Morris, S.}{2005}{Robust mechanism design}{\emet 73}{1771-1813}

\paper{Bergemann, D., Morris, S.}{2013}{An introduction to robust mechanism design}{{\it Foundations and Trends in Microeconomics\/} 8}{169-230}

\paper{\v Copi\v c, J., Ponsat\'\i, C.}{2016}{Optimal robust bilateral trade: Risk neutrality}{\jet 163}{276-287}

\paper{Cripps, M., Swinkels, J.}{2006}{Efficiency of large double auctions}{\emet 74}{47-92}

\paper{Drexl, M., Kleiner, A.}{2015}{Optimal private good allocation: The case for a balanced budget}{\geb 94}{169-181}

\paper{Hagerty, K., Rogerson, W.}{1987}{Robust trading mechanisms}{\jet 42}{94-107}

\paper{Kojima, F., Yamashita, T.}{2017}{Double auctions with interdependent values: Incentives and efficiency}{\te 12}{1393-1438}

\paper{Kuzmics, C., Steg, J.-H.}{2017}{On public good provision mechanisms with dominant strategies and balanced budget}{\jet 170}{56-69}

\paper{McAfee, P.}{1992}{A dominant strategy double auction}{\jet 56}{434-450}

\paper{Miyagawa, E.}{2001}{House allocation with transfers}{\jet 100}{329-355}

\paper{Myerson, R.}{1981}{Optimal auction design}{{\it Mathematics of Operations Research\/} 6}{58-73}

\paper{Myerson, R., Satterthwaite, M.}{1983}{Efficient mechanisms for bilateral trading}{\jet 29}{265-281}

\paper{Rustichini, A., Satterthwaite, M., Williams, S.}{1994}{Convergence to efficiency in a simple market with incomplete information}{\emet 62}{1041-1063}

\paper{Satterthwaite, M., Sonnenschein, H.}{1981}{Strategy-proof allocation mechanisms at differentiable points}{\res 48}{587-597}

\paper{Satterthwaite, M., Williams, S.}{2002}{The optimality of a simple market mechanism}{\emet 70}{1841-1863}

\paper{Satterthwaite, M., Williams, S., Zachariadis, K.}{2014}{Optimality versus practicality in market design: A comparison of two double auctions}{\geb 86}{248-263}

\paper{Shao, R., Zhou, L.}{2016}{Optimal allocation of an indivisible good}{\geb 100}{95-112}

\paper{Thomson, W.}{2016}{Non-bossiness}{{\it Social Choice and Welfare\/} 47}{665-696}

\paper{Wilson, R.}{1985}{Incentive efficiency of double auctions}{\emet 53}{1101-1115}

\paper{Yoon, K.}{2001}{The modified Vickrey double auction}{\jet 101}{572-584}

\bye